\documentclass[aps,prl,reprint,nofootinbib]{revtex4-2}
\pdfoutput=1

\usepackage{amsmath,amssymb,bm}
\usepackage{mathtools}

\newcommand{\C}{\mathbb C}
\newcommand{\R}{\mathbb R}
\newcommand{\cP}{\mathcal P}

\newcommand{\cG}{\mathcal G}

\begin{document}

\title{General Relativity in Three Lines}

\author{Pawe{\l} Nurowski}
\affiliation{Center for Theoretical Physics, Polish Academy of Sciences, Warsaw, Poland}

\begin{abstract}
We show that complex general relativity can be formulated directly
in terms of three affine lines assigned to each spacetime point.
The lines are represented by Pl\"ucker bivectors, and Urbantke's
construction applied to the associated two-forms reconstructs the
conformal spacetime metric.  Starting from the chiral Pleba\'nski
action, the simplicity constraints turn the three two-forms into a
null internal basis of simple bivectors and thereby into three
projective lines.  A $3+1$ decomposition then yields a first-order
canonical theory whose configuration variable is the weighted triple
of lines and whose conjugate variable is a line momentum.  The
resulting constraint system consists of three internal Gauss
constraints, three vector constraints, and one Hamiltonian
constraint.

A notable feature of the line formulation is that the equation
obtained by varying the line momentum is pointwise algebraic after
the spatial integrations by parts.  Consequently the momentum can
be eliminated explicitly wherever the corresponding Legendre map is
invertible, giving a second-order action whose only dynamical
configuration variable is the weighted triple of lines.  The line
curvature separates naturally into a cosmological trace and the Weyl
matrix of the selected chiral sector. The Legendre map becomes singular on a distinguished null-evolution
locus where the spacetime metric remains nondegenerate.  There its
rank drops from nine to six, giving three algebraic compatibility
conditions.  We construct a finite invariant partial Legendre
transform and a smooth constrained second-order continuation through
this locus.  This analysis provides a regular framework for regimes
in which the evolution direction becomes null and may be useful in
the study of near-horizon geometries and related gravitational
phenomena.
\end{abstract}

\maketitle
\section*{Introduction}

The geometric idea underlying this paper is elementary.  Three
mutually skew straight lines in real affine three-space determine a
doubly ruled quadric.  One way to see this is the classical
construction described by Hilbert and Cohn--Vossen
\cite[p.14]{HilbertCohnVossen}.  Choose a point $P$ on the first line
$\ell_1$.  There is a unique plane containing $P$ and the second
line $\ell_2$, and another unique plane containing $P$ and the third
line $\ell_3$.  The intersection of these two planes is a line
through $P$ meeting all three original lines.  As $P$ moves along
$\ell_1$, these transversals sweep out one ruling of a smooth
quadric, while $\ell_1,\ell_2,\ell_3$ belong to the other ruling.
Thus three skew lines determine the entire ruled quadric.  No
notion of distance, angle, or metric enters this construction.

The relation with a four-dimensional metric appears after passing
to projective space.  Let $V_{\mathbb R}$ be a real
four-dimensional vector space, so that
$\mathbb P(V_{\mathbb R})\simeq\mathbb{RP}^3$.  The projective
completion of the ruled quadric is the zero set of a nondegenerate
homogeneous quadratic form,
\[
 Q=\bigl\{[X]\in\mathbb P(V_{\mathbb R}):
              q(X,X)=0\bigr\}.
\]
Equivalently, before projectivization one has the cone
\[
 {\cal N}_q
 :=\bigl\{X\in V_{\mathbb R}\setminus\{0\}:q(X,X)=0\bigr\}.
\]
Multiplying $q$ by a nonzero constant leaves $Q$ unchanged.
Consequently the three lines determine not a preferred quadratic
form but its conformal class $[q]$.

The signature of this real quadratic form is already fixed by the
ruled geometry.  A projective line $\ell=\mathbb P(L)$ contained in
$Q$ comes from a two-dimensional subspace
$L\subset V_{\mathbb R}$ on which $q$ vanishes identically.  Thus
$L$ is a totally null two-plane.  If the signature of a
nondegenerate quadratic form on a four-dimensional real vector
space is $(p,r)$, the dimension of a totally null subspace cannot
exceed $\min(p,r)$.  Since here such a subspace has dimension two,
one must have
\[
 p\geq2,\qquad r\geq2,\qquad p+r=4,
\]
and hence
\[
 \boxed{\operatorname{sign}(q)=(2,2).}
\]
The two rulings of the quadric are precisely the two families of
projectivized maximal totally null two-planes of this split
conformal structure.  In this sense the elementary construction can
be summarized as
\[
 \boxed{
 \text{three skew real lines}
 \ \longrightarrow\
 Q\subset\mathbb{RP}^3
 \ \longrightarrow\
 [q]_{(2,2)} .
 }
\]

This observation suggests the pointwise principle developed in the
present paper.  Let $M$ be a four-manifold.  At a point $x\in M$ we
replace the fixed four-dimensional vector space above by the tangent
space $T_xM$, or, for the chiral theory, by its complexification
$T_xM\otimes\mathbb C$.  We then regard three projective lines as
the primary geometric data and interpret the quadric which they
determine as the projectivized null cone of a conformal metric at
$x$:
\[
 \boxed{
 \{\ell_1(x),\ell_2(x),\ell_3(x)\}
 \subset\mathbb P(T_xM\otimes\mathbb C)
 \quad\Longrightarrow\quad
 [g_x].
 }
\]
Thus the usual logical direction is reversed.  Rather than starting
with a spacetime metric and asking for its null geometry, we start
with three lines and reconstruct the conformal metric from them.

The real three-line picture above naturally gives split signature.
General relativity in its chiral formulation is most naturally
treated over $\mathbb C$, and we therefore complexify this
construction.  Different real forms of the resulting complex
geometry are then selected by reality conditions.  The purpose of
the paper is to make this pointwise idea explicit and dynamical.
We first express the conformal metric directly in terms of the
Pl\"ucker data of the three lines and identify this construction
with Urbantke's formula.  We then show that the simplicity
constraints of the chiral Pleba\'nski theory turn its three
two-forms precisely into such line data, and we develop the
resulting first-order canonical theory, constraint algebra and
second-order line action.

\section{Geometry of three lines}

Let $E\simeq\C^3$ be an affine three-space.  Throughout, Greek
letters used as spacetime indices range over $0,1,2,3$, spatial
coordinate indices $a,b,c,\ldots$ range over $1,2,3$, and internal
indices $I,J,K,\ldots$ and $i,j,k,r,s,\ldots$ range over $1,2,3$.
Repeated indices in tensor contractions are summed unless explicitly
stated otherwise.  Whenever an index labels a componentwise relation
rather than a contraction, we indicate this by ``no sum''.  We use
the numerical alternating symbols $\epsilon_{ijk}$ and
$\epsilon^{ijk}$, normalized by
$\epsilon_{123}=\epsilon^{123}=+1$; the placement of their indices
is not obtained by raising or lowering with $h_{ij}$.  The symbol
$\operatorname{Tr}$ denotes the ordinary matrix trace, and for
matrices we write $[A,B]:=AB-BA$.  Superscript $T$ denotes
transpose, an overbar complex conjugation, and $\dagger$ Hermitian
transpose.  We write $\mathrm i^2=-1$ for the imaginary unit.

At each point $x$ of the four-manifold $M$ consider three affine
lines in $E$,
\begin{equation}
 \ell_i(x)=\bm p_i(x)+\lambda_{\rm aff}\,\bm v_i(x),
 \qquad \lambda_{\rm aff}\in\C,\qquad i=1,2,3.
 \label{eq:lines}
\end{equation}
These are the complex pointwise counterparts of the three real skew
lines in the introductory construction.

The four-dimensional space enters by homogenization:
$V=\C e_0\oplus E\simeq\C^4$. At each $x\in M$ we identify $V$ with
the complexified tangent space $T_xM\otimes\C$.  Under this
identification the bilinear form reconstructed below from the three
lines is a conformal spacetime metric at $x$.  The projective
completion of $\ell_i$ is the two-plane with Pl\"ucker bivector
\begin{equation}
 \bm{\cP}_i=(e_0+\bm p_i)\wedge\bm v_i\in\Lambda^2V,
 \qquad i=1,2,3
 \quad\text{(no sum over $i$)}.
 \label{eq:plucker}
\end{equation}
The base point $\bm p_i$ is not intrinsic, since, for each fixed
$i$,
\[
 \bm p_i\mapsto\bm p_i+s_i\bm v_i \qquad\text{(no sum over $i$)}
\]
leaves the line unchanged.

From now on $\bm a\cdot\bm b$ and $\bm a\times\bm b$ denote the
complex-bilinear extensions to $\C^3$ of the ordinary scalar and
vector products on $\R^3$.  It is therefore convenient to introduce
the Pl\"ucker moment, for each $i=1,2,3$,
\begin{equation}
 \bm m_i:=\bm p_i\times\bm v_i,\qquad
 \bm m_i\cdot\bm v_i=0
 \qquad\text{(no sum over $i$)}.
 \label{eq:moment}
\end{equation}
Conversely, for each fixed $i$ (no sum over $i$), every
$\bm m_i$ satisfying
\[
 \bm m_i\cdot\bm v_i=0
 \qquad\text{(no sum over $i$)}
\]
can be written as
\[
 \bm m_i=\bm p_i\times\bm v_i
 \qquad\text{(no sum over $i$)},
\]
with $\bm p_i$ unique modulo
\[
\bm p_i\mapsto\bm p_i+s_i\bm v_i
\qquad\text{(no sum over $i$)}.
\]
Thus $(\bm m_i,\bm v_i)$ contains exactly the same line data as
$(\bm p_i,\bm v_i)$ for each fixed $i$.

For a vector $X\in V$ and a $k$-form $\alpha$ we use the interior
product
\begin{equation}
 (\iota_X\alpha)(Y_1,\ldots,Y_{k-1})
 =\alpha(X,Y_1,\ldots,Y_{k-1}),
\end{equation}
and, for a decomposable bivector $u\wedge v$, we set
$\iota_{u\wedge v}:=\iota_v\iota_u$.  Fix a nonzero
$\Omega\in\Lambda^4V^*$.  Then $\iota_{\bm{\cP}_i}\Omega$ is a
two-form and $\iota_X\iota_{\bm{\cP}_i}\Omega$ a one-form.  Applying
Urbantke's construction~\cite{Urbantke1984} to the three two-forms
$\iota_{\bm{\cP}_i}\Omega$, the three lines define the
four-form-valued symmetric bilinear form
\begin{equation}
 \cG(X,Y)=\frac16\epsilon_{ijk}
 (\iota_X\iota_{\bm{\cP}_i}\Omega)\wedge
 (\iota_Y\iota_{\bm{\cP}_j}\Omega)\wedge
 (\iota_{\bm{\cP}_k}\Omega) ,
 \label{eq:urb}
\end{equation}
Changing $\Omega$ or
rescaling the Pl\"ucker representatives changes only the overall
conformal factor.  The object $\cG$ is therefore the intrinsic,
four-form-valued version of the conformal metric.  After fixing
$\Omega$, we define the ordinary symmetric bilinear representative
$g$ by
\begin{equation}
        \cG(X,Y)=g(X,Y)\,\Omega .
        \label{eq:Gtog}
\end{equation}
In a chosen affine coframe $(dt,dx^a)$ we denote the matrix of $g$ by
$G[\bm m,\bm v]=(G_{\mu\nu})$.  Finally, the notation $ds^2$ will mean
nothing more than the quadratic-form notation for this same
representative,
\begin{equation}
        ds^2:=G_{\mu\nu}\,dx^\mu dx^\nu .
        \label{eq:Gtods}
\end{equation}
Thus $\cG$, $G$, and $ds^2$ below are three presentations of the same
conformal metric: intrinsic, matrix, and line-element form,
respectively.

Equations~\eqref{eq:urb}--\eqref{eq:Gtog} give
\begin{equation}
 G[\bm m,\bm v]=
 \begin{pmatrix}
 A&\bm q^{\,T}\\[1mm] \bm q&H
 \end{pmatrix},
 \label{eq:block}
\end{equation}
where all entries are explicit functions of the equivalent
Pl\"ucker pairs $(\bm m_i,\bm v_i)$; $m_{ia}$ denotes the components
of $\bm m_i$:
\begin{align}
 A={}&\bm m_1\cdot(\bm m_2\times\bm m_3),
 \label{eq:A}\\[1mm]
 \bm q={}&\frac12\Big[
 (\bm m_2\times\bm m_3)\times\bm v_1
 +(\bm m_3\times\bm m_1)\times\bm v_2
 \nonumber\\
 &\hspace{18mm}
 +(\bm m_1\times\bm m_2)\times\bm v_3
 \Big],
 \label{eq:q}\\[1mm]
 H_{ab}={}&\frac12\sum_{\mathrm{cyc}(i,j,k)}
 \Big[
 (\bm v_j\times\bm v_k)_a\,m_{ib}
 +m_{ia}\,(\bm v_j\times\bm v_k)_b
 \Big].
 \label{eq:H}
\end{align}
Here $\sum_{\mathrm{cyc}(i,j,k)}$ denotes the sum over
$(i,j,k)=(1,2,3),(2,3,1),(3,1,2)$.

Equivalently, the same representative in line-element notation is
\begin{equation}
 ds^2=A\,dt^2+2q_a\,dt\,dx^a+H_{ab}\,dx^a dx^b .
 \label{eq:lineelement}
\end{equation}
Equations~\eqref{eq:A}--\eqref{eq:H} therefore make the
four-dimensional conformal metric an explicit algebraic function of
three affine lines, with no metric structure assumed on $E$.

The representation has the elementary redundancies
\begin{align}
 &\bm p_i\mapsto\bm p_i+s_i\bm v_i,\qquad
 \bm v_i\mapsto c_i\bm v_i,\\& c_i\neq0,
 \qquad i=1,2,3,
 \quad\text{(no sum over $i$)}.\nonumber
 \label{eq:gauge}
\end{align}
For each fixed $i$, the first transformation leaves
$\bm m_i$ unchanged, while the second sends
$(\bm m_i,\bm v_i)\mapsto c_i(\bm m_i,\bm v_i)$ and rescales its
Pl\"ucker bivector; consequently
$G\mapsto(c_1c_2c_3)G$ and the conformal structure is unchanged.

On the open set $A\neq0$ define
\begin{equation}
 \theta^0=dt+\beta_a dx^a,\qquad
 \beta=\frac qA,\qquad
 K_{\rm sp}=\frac{AH-qq^T}{A^2}.
 \label{eq:tauK}
\end{equation}
Then the same line element, rescaled by the nonzero conformal
factor $A^{-1}$, takes the orthogonal form
\begin{equation}
 \frac1A\,ds^2=(\theta^0)^2+(K_{\rm sp})_{ab}\,dx^a dx^b .
 \label{eq:orth}
\end{equation}
The conformal class contains a real metric precisely when
\begin{equation}
 \boxed{\;
 \bar A\,q_a-A\,\bar q_a=0,\qquad
 \bar A\,H_{ab}-A\,\bar H_{ab}=0 .
 \;}
 \label{eq:reality}
\end{equation}
so that $\beta$ and the symmetric matrix $K_{\rm sp}$ are real.
For a nondegenerate configuration its signature is then read off
entirely from $K_{\rm sp}$:
\begin{equation}
 \begin{array}{rcl}
 K_{\rm sp}>0
   &\Longleftrightarrow& (4,0),\\[1mm]
 \det K_{\rm sp}<0
   &\Longleftrightarrow& (1,3)\ {\rm or}\ (3,1),\\[1mm]
 \det K_{\rm sp}>0,\quad K_{\rm sp}\not>0
   &\Longleftrightarrow& (2,2).
 \end{array}
 \label{eq:signatures}
\end{equation}
In particular three real skew lines give the split real form, whereas
the Euclidean and Lorentzian metrics arise as different real sections
of the same complex line geometry.

For the projective action it is simplest to return temporarily to
the affine representatives $(\bm p_i,\bm v_i)$ of
Eq.~\eqref{eq:lines}.  The transformations preserving a fixed
conformal structure are most transparent in homogeneous coordinates.

For $a_0\in\C$, $\bm b,\bm c\in\C^3$ and
$\mathsf D\in\operatorname{Mat}_{3\times3}(\C)$, write
\begin{equation}
 T=\begin{pmatrix}a_0&\bm b^T\\ \bm c&\mathsf D\end{pmatrix}.
\end{equation}
Its induced action on an affine line is
\begin{align}
 \bm p_i'&=
 \frac{\bm c+\mathsf D\bm p_i}
      {a_0+\bm b\!\cdot\bm p_i},
 \nonumber\\
 \bm v_i'&=
 (a_0+\bm b\!\cdot\bm p_i)\mathsf D\bm v_i
 -(\bm b\!\cdot\bm v_i)
  (\bm c+\mathsf D\bm p_i),
 \label{eq:pvaction}
\end{align}
with  $i=1,2,3$ and there is no sum over $i$.
A transformation is a conformal symmetry of the reconstructed
metric iff, for some $\varpi\in\C^*$,
\begin{equation}
 T^TGT=\varpi G .
 \label{eq:confstab}
\end{equation}
To make the complex orthogonal action explicit, identify
$V\simeq M_2(\C)$ as follows and let $G_0$ denote the matrix of the
determinant quadratic form:
\begin{align}
 &x=(x^0,x^1,x^2,x^3)^T
 \ \longleftrightarrow\
 X=\begin{pmatrix}x^0&x^1\\x^2&x^3\end{pmatrix},\\
 &x^T G_0x=\det X=x^0x^3-x^1x^2.
 \label{eq:segreid}
\end{align}
For $S_+,S_-\in SL(2,\C)$ consider the spin action
$X\mapsto S_+XS_-^T$.  Via \eqref{eq:segreid} it induces
$T=T(S_+,S_-)$ on $V$:
\begin{equation}
 x\mapsto T(S_+,S_-)x .
 \label{eq:spinT}
\end{equation}
Since $\det S_+=\det S_-=1$,
\begin{equation}
 T(S_+,S_-)^T G_0T(S_+,S_-)=G_0 .
\end{equation}
Thus
\[
 SO(4,\C)\simeq
 [SL(2,\C)\times SL(2,\C)]/\mathbb Z_2 .
\]
For a general representative
$G=C_0^TG_0C_0$, with $C_0\in GL(4,\C)$, its orthogonal
transformations are
$C_0^{-1}T(S_+,S_-)C_0$.  The relevant real forms are
\begin{equation}
 \begin{array}{rcl}
 (2,2)&:&S_+,S_-\in SL(2,\R),\\
 (4,0)&:&S_+,S_-\in SU(2),\\
 (1,3)&:&S_+\in SL(2,\C),\quad S_-=\bar S_+ .
 \end{array}
 \label{eq:realforms}
\end{equation}
In the Lorentzian case this gives
$X\mapsto S_+XS_+^\dagger$.
The projectivized null cone
\begin{equation}
 Q_G=\{[x]\in\mathbb P(V):x^TGx=0\}
 \label{eq:nullquadric}
\end{equation}
is a smooth quadric surface.  For $G_0$ it is simply
$\det X=0$, hence every point can be written $X=u\,w^T$ and
$Q_{G_0}\simeq\mathbb P^1_u\times\mathbb P^1_w$.  Fixing $[u]$ and
varying $[w]$ gives one family (ruling) of projective lines; fixing
$[w]$ gives the other.  Under $X\mapsto S_+XS_-^T$, the factor $S_+$ moves
the lines of the first ruling, $[u]\mapsto[S_+u]$, whereas $S_-$ acts
along each such line, $[w]\mapsto[S_-w]$; for the opposite ruling their
roles are interchanged.  Conjugation by $C_0$ gives the same statement
for the quadric $Q_G$.

\subsection{Example: A known family from three lines.}
In coordinates $(p,q,\sigma,\tau)$ let
\begin{align}
 P&=P(p),\qquad Q=Q(q),\nonumber\\
 F&=1+p^2q^2,\qquad \Delta=\sqrt{PQ}.
\end{align}
Here $\mathrm i^2=-1$ is the imaginary unit.  Consider the three
lines \eqref{eq:lines} with
\begin{align}
 \bm p_1&=(-Q,0,-q^2),&
 \bm v_1&=(Qp^2,-\mathrm i P,F),\nonumber\\
 \bm p_2&=(Q,0,-q^2),&
 \bm v_2&=(-Qp^2,\mathrm i P,F),\nonumber\\
 \bm p_3&=(0,\Delta,-q^2),&
 \bm v_3&=\left(-\mathrm i,-p^2,\frac{F}{\Delta}\right).
 \label{eq:PDlines}
\end{align}
Forming
$\bm m_i=\bm p_i\times\bm v_i$ for $i=1,2,3$
(no sum over $i$), and substituting the resulting Pl\"ucker pairs into Eqs.~\eqref{eq:A}--\eqref{eq:H}
determines the matrix $G[\bm m,\bm v]$.  In the equivalent
line-element notation \eqref{eq:Gtods},
\begin{align}
 \frac{ds^2[\bm m,\bm v]}{-2\mathrm i\,PQF}
 ={}&\frac{F}{P}\,dp^2
 +\frac{P}{F}(d\sigma+q^2d\tau)^2 \nonumber\\
 &+\frac{F}{Q}\,dq^2
 -\frac{Q}{F}(d\tau-p^2d\sigma)^2 .
 \label{eq:PDmetric}
\end{align}
For the quartic functions $P(p)$ and $Q(q)$ fixed by the
Einstein--Maxwell equations, Eq.~\eqref{eq:PDmetric} is precisely the
general Pleba\'nski--Demia\'nski family \cite{PlebanskiDemianski}.

\section{Lagrangian formulation and Einstein equations}

We now endow the three-line geometry with dynamics.  Let $e_I$ be
a basis of $\C^3$ and let
$\eta_{IJ}=\operatorname{diag}(1,1,1)$ be its fixed complex-bilinear
internal metric, with inverse $\eta^{IJ}$.  Let
$\epsilon^I{}_{JK}$ be the standard antisymmetric symbol.  Let
$B=B^I e_I$ be a $\C^3$-valued two-form and
${\cal A}={\cal A}^I e_I$ a chiral connection, whose curvature is
${\cal F}^I:=d{\cal A}^I+
\frac12\epsilon^I{}_{JK}{\cal A}^J\wedge{\cal A}^K$.
Let $\Psi^{IJ}$ be symmetric and $\eta$-tracefree, let $\Lambda$ be
the cosmological constant, and let $\kappa$ denote the gravitational
coupling.  The complex chiral Pleba\'nski action~\cite{Plebanski1977} is
\begin{equation}
 S_{\rm P}=\frac1\kappa\int_M\left[
 B^I\wedge{\cal F}^I
 -\frac12(\Psi^{IJ}+\Lambda\eta^{IJ})B^I\wedge B^J
 \right].
 \label{eq:Plebanski}
\end{equation}
The canonical connection formulation of general relativity based on
self-dual variables was introduced by Ashtekar
\cite{Ashtekar1986,Ashtekar1987}.  Manifestly covariant Lagrangian
origins of the same canonical formulation were exhibited shortly
afterwards by Samuel~\cite{Samuel1987} and by Jacobson and
Smolin~\cite{JacobsonSmolin1988}.  The relation with the
Pleba\'nski self-dual two-form formulation was subsequently made
particularly explicit by Capovilla, Jacobson, Dell and
Mason~\cite{CapovillaJacobsonDellMason1991}.  Our purpose here is not
to introduce another connection formulation, but to rewrite this
chiral dynamics directly in the geometry of the three lines.

Define the four-form
\[
 {\cal V}_4:=\frac13\eta_{KL}B^K\wedge B^L .
\]
Variation with respect to $\Psi^{IJ}$ gives the five simplicity
equations
\begin{equation}
 B^I\wedge B^J={\cal V}_4\,\eta^{IJ}.
 \label{eq:simplicityorth}
\end{equation}
Thus the six symmetric products $B^I\wedge B^J$ are all fixed by
the single four-form ${\cal V}_4$.

To expose their line geometry, change only the \emph{internal} basis.
Choose three null vectors $n_i\in\C^3$ satisfying
\begin{equation}
 h_{ij}:=\eta_{IJ}n_i{}^I n_j{}^J=
 \begin{pmatrix}
 0&1&1\\
 1&0&1\\
 1&1&0
 \end{pmatrix}.
 \label{eq:nullinternal}
\end{equation}
Let $(h^{ij}):=(h_{ij})^{-1}$ and define $n^i:=h^{ij}n_j$.  Then
expand the same field as
\begin{equation}
 B=B_i\,n^i,\qquad
 B_i=\eta_{IJ}n_i{}^I B^J .
 \label{eq:Bbasischange}
\end{equation}
No new fields have been introduced: the $B_i$ are simply the
components of the original $B^I$ in the null internal basis.
Equation~\eqref{eq:simplicityorth} then becomes
\begin{equation}
 \boxed{\;B_i\wedge B_j={\cal V}_4\,h_{ij}.\;}
 \label{eq:simplicity}
\end{equation}
For each fixed $i=1,2,3$,
\[
 h_{ii}=0
 \qquad\text{(no sum over $i$)},
\]
and therefore
\[
 B_i\wedge B_i=0
 \qquad\text{(no sum over $i$)}.
\]
In four dimensions a nonzero two-form is decomposable iff its square
vanishes; hence, on the nondegenerate sector ${\cal V}_4\neq0$,
there exist complex one-forms $\alpha_i,\beta_i\in\Omega^1(M,\C)$
such that
\[
 B_i=\alpha_i\wedge\beta_i,
 \qquad i=1,2,3
 \quad\text{(no sum over $i$)}.
\]
Equivalently, after choosing a nonzero four-form $\Omega$, there is a
decomposable bivector ${\cal P}_i$ such that
\begin{equation}
 B_i=\iota_{{\cal P}_i}\Omega .
 \label{eq:BPlucker}
\end{equation}
A decomposable bivector determines a projective two-plane in
$\C^4$, i.e. one projective line.  Thus the diagonal simplicity
conditions make each $B_i$ decomposable and hence define three
Pl\"ucker lines, while the off-diagonal conditions constrain their
mutual position.

Choose now coordinates $(t,x^a)$ adapted to the affine splitting
of Sec.~I, let $\Sigma$ denote a hypersurface $t=\mathrm{const}$,
and write
\[
 \Omega=dt\wedge\epsilon,\qquad
 \epsilon=\rho\,d^3x,\qquad
 d^3x:=dx^1\wedge dx^2\wedge dx^3,
\]
with nonzero coordinate density $\rho$.  We now suppress boldface
for the spatial vectors of Sec.~I. For each fixed $i=1,2,3$, $B_i$ determines uniquely a spatial
one-form $\mu_i$ and a spatial vector $v_i$:
\begin{equation}
 B_i=dt\wedge\mu_i+\iota_{v_i}\epsilon,\qquad
 L_i{}^a:=\rho\,v_i^a,
 \qquad i=1,2,3.
 \label{eq:Bsplit}
\end{equation}
The one-form $\mu_i$ corresponds to the Pl\"ucker moment
$\bm m_i=\bm p_i\times\bm v_i$ of Sec.~I.  Explicitly,
\begin{equation}
 \mu_i:=\iota_{v_i}\iota_{p_i}\epsilon
      =\rho\,m_{ia}\,dx^a,
 \qquad i=1,2,3.
 \label{eq:mu-moment}
\end{equation}
Consequently,
\begin{equation}
 \mu_i(v_i)
 =\rho\,\bm m_i\!\cdot\!\bm v_i=0,\qquad
 B_i\wedge B_i
 =2\,\mu_i(v_i)\,dt\wedge\epsilon=0,
 \label{eq:mu-identities}
\end{equation}
with $i=1,2,3$ and no sum over $i$.

Hence $\mu_i$ carries exactly the same moment data as $\bm m_i$,
now in covector form, while $L_i{}^a$ is the weighted line
direction.  We regard $L=(L_i{}^a)$ as a $3\times3$ matrix and from
now on work in the nondegenerate sector
\[
 D:=\det L\neq0 .
\]

Writing
\[
 {\cal V}_4=\nu\,dt\wedge\epsilon,
\]
the off-diagonal simplicity equations are
\[
 \mu_i(v_j)+\mu_j(v_i)=\nu h_{ij}.
\]
For independent $v_i$ they are parametrized by a common translation
$N$ and
\[
 \lambda=\frac{\nu}{2\epsilon(v_1,v_2,v_3)}.
\]
Set
\[
 L_i{}^a=\rho v_i^a,\qquad
 \omega=\frac{\lambda}{\rho}.
\]
Because $L$ is invertible, define $s^i$ by
\[
 N^a=s^iL_i{}^a .
\]
Then the affine lines are
\begin{align}
 \ell_1&=-N+\omega(L_3-L_2)+\C L_1,\nonumber\\
 \ell_2&=-N+\omega(L_1-L_3)+\C L_2,\nonumber\\
 \ell_3&=-N+\omega(L_2-L_1)+\C L_3 .
 \label{eq:dynamiclines}
\end{align}
Thus $s^i$ are the line-frame components of the common translation,
while $\omega$ controls the opening.

Let $\{n_i\}$ be the internal null basis introduced above.  Expand
the Pleba\'nski connection in this basis and in the coordinates
$x^\mu=(t,x^a)$ as
\begin{equation}
 {\cal A}={\cal A}^i n_i
 =\left({\cal A}_0{}^i\,dt+a_a{}^i\,dx^a\right)n_i .
 \label{eq:connection-split}
\end{equation}
Thus $a_a{}^i$ are the spatial components of the chiral
connection.  Writing ${\cal F}={\cal F}^i n_i$, the BF term is
$B_i\wedge{\cal F}^i$.  With a dot denoting $\partial_t$, the
curvature contains
\[
 {\cal F}^i
 =dt\wedge\left(\dot a_a{}^i-\cdots\right)dx^a
   +\hbox{terms without }\dot a .
\]
Using $B_i=dt\wedge\mu_i+\iota_{v_i}\epsilon$ and
$L_i{}^a=\rho v_i^a$, the part of the Pleba\'nski action containing
time derivatives is therefore
\begin{equation}
 S_{\rm P}\big|_{\rm kin}
 =\frac1\kappa\int dt\,d^3x\,
 L_i{}^a\dot a_a{}^i .
 \label{eq:kinetic}
\end{equation}
Hence the corresponding symplectic potential (a one-form on field
space, with $\delta$ denoting field variation) is
\begin{equation}
 \Theta_{\rm old}
 =\frac1\kappa\int_\Sigma
 L_i{}^a\,\delta a_a{}^i .
 \label{eq:theta-old}
\end{equation}
Its field-space exterior derivative
$\Omega_{\rm symp}:=\delta\Theta_{\rm old}$ is the symplectic form.

We want the line variables $L_i{}^a$, rather than the connection
$a_a{}^i$, to be the configuration variables.  Since
\[
 \delta(L_i{}^a a_a{}^i)
 =a_a{}^i\,\delta L_i{}^a
 +L_i{}^a\,\delta a_a{}^i ,
\]
we may subtract the exact field-space variation
$\delta[\kappa^{-1}\int_\Sigma L_i{}^a a_a{}^i]$, which leaves
$\Omega_{\rm symp}$ unchanged.  Thus we use instead
$\Theta_{\rm line}:=-\kappa^{-1}\int_\Sigma
a_a{}^i\,\delta L_i{}^a$.  Defining
\begin{equation}
 \pi_a{}^i:=-a_a{}^i,\qquad
 \Pi_j{}^i:=L_j{}^a\pi_a{}^i
          =-L_j{}^a a_a{}^i ,
 \label{eq:Pi-definition}
\end{equation}
the symplectic potential can be written equivalently as
\begin{equation}
 \boxed{
 \Theta_{\rm line}
 =\frac1\kappa\int_\Sigma
 \operatorname{Tr}\!\left(\Pi\,\delta L\,L^{-1}\right).
 }
 \label{eq:symplectic}
\end{equation}
Thus $\pi_a{}^i$ is the Darboux momentum conjugate to
$L_i{}^a$, whereas $\Pi$ is its right-trivialized form and pairs
canonically with the Lie-algebra-valued variation
$\Xi:=\delta L\,L^{-1}$.  Consequently, $(L,\Pi)$ is a
right-trivialized $T^*GL(3,\C)$ phase space.  Correspondingly define
the right-trivialized velocity $U:=\dot L\,L^{-1}$, so that the
kinetic term is $\operatorname{Tr}(\Pi U)$.

Set $R:=L^{-1}$.  Since $L_i{}^a$ is a contravariant spatial vector
density of weight
$+1$, the already defined $D=\det L$ has weight $+2$, while $R$ has
weight $-1$.  Let $\delta_i{}^j$ denote the Kronecker symbol.
Now define
\begin{align}
 \vartheta_i&:=L_i{}^a\partial_a\log D,\nonumber\\
 \Gamma_{ij}{}^k
 &:=
 R_a{}^k
 \left(L_i{}^b\partial_bL_j{}^a-L_j{}^b\partial_bL_i{}^a\right)
 -\frac12\vartheta_i\delta_j{}^k
 +\frac12\vartheta_j\delta_i{}^k ,
 \label{eq:anhol}
\end{align}
and $\gamma_i:=\Gamma_{ji}{}^j=\partial_aL_i{}^a$.
For a spatial density $X$ of weight $w$ set
\begin{equation}
 \nabla_i^{(w)}X
 :=L_i{}^a\left(\partial_aX-\frac w2
 (\partial_a\log D)X\right).
 \label{eq:weightedder}
\end{equation}
The Lie bracket of the internal null basis is
\begin{equation}
 [n_i,n_j]=f_{ij}{}^k n_k,
 \qquad
 f_{ij}{}^k:=\sqrt2\,\epsilon_{ijm}h^{mk}.
 \label{eq:nullstructure}
\end{equation}
Thus $f_{ij}{}^k$ are the structure constants used below.

The intrinsic line
curvature and its dual are
\begin{align}
 {\cal F}_{ij}{}^k={}&
 -\nabla_i^{(1)}\Pi_j{}^k+\nabla_j^{(1)}\Pi_i{}^k
 \nonumber\\
 &+\Gamma_{ij}{}^m\Pi_m{}^k
 +f_{mn}{}^k\Pi_i{}^m\Pi_j{}^n,
 \nonumber\\
 {\cal R}^{ij}:={}&\frac12\epsilon^{ikl}{\cal F}_{kl}{}^j .
 \label{eq:linecurvature}
\end{align}
With
\begin{equation}
 \widehat\Pi^{ij}:=
 h^{ik}\Pi_k{}^j+\frac1{2\sqrt2}\epsilon^{kij}\gamma_k ,
 \label{eq:shiftedPi}
\end{equation}
using $X^{[ij]}:=\frac12(X^{ij}-X^{ji})$, the seven Einstein
constraints take the $3+3+1$ form
\begin{equation}
 \boxed{\;
 \widehat\Pi^{[ij]}=0,\qquad
 {\cal R}^{[ij]}=0,\qquad
 h_{ij}{\cal R}^{ij}=3\Lambda D .
 \;}
 \label{eq:sevenconstraints}
\end{equation}
Equivalently,
\begin{equation}
 {\cal G}_i:=\gamma_i-f_{ij}{}^k\Pi_k{}^j,\quad
 {\cal V}_i:={\cal F}_{ij}{}^j,\quad
 {\cal H}:=h_{ij}{\cal R}^{ij}-3\Lambda D ,
 \label{eq:componentconstraints}
\end{equation}
since
$\widehat\Pi^{[ij]}=(2\sqrt2)^{-1}\epsilon^{kij}{\cal G}_k$
and
${\cal R}^{[ij]}=-\frac12\epsilon^{kij}{\cal V}_k$.

Thus \eqref{eq:sevenconstraints} gives the complete first-class
constraint sector of the chiral theory.  The three conditions
$\widehat\Pi^{[ij]}=0$ are the internal Gauss constraints, the three
conditions ${\cal R}^{[ij]}=0$ are the vector constraints, and the
remaining $h$-trace condition is the scalar (Hamiltonian)
constraint of general relativity. As shown below,
the raw vector constraint generates a spatial diffeomorphism only
up to a field-dependent internal rotation; the improved generator
\eqref{eq:improvedD} generates the ordinary spatial Lie derivative.
The vector and scalar constraints restrict the gravitational
initial data on $\Sigma$, while the complementary spacetime
Einstein equations occur among the evolution equations following
from the first-order action.  Notice in particular that the vector
and scalar constraints fix the antisymmetric and $h$-trace parts
of ${\cal R}^{ij}$, whereas its symmetric $h$-tracefree part
remains and will be identified below with the chosen chiral Weyl
curvature.

Let $\zeta^i$ be density-weight-zero multipliers generating
internal $SO(h,\C)$ rotations, where
\[
 SO(h,\C):=
 \{S_h\in SL(3,\C):S_h^T h S_h=h\}.
\]
The translation components $s^i$ and the opening variable $\omega$
have density weight $-1$. The non-kinetic Pleba\'nski terms then
reduce to the three constraint terms, so the complete action is
\begin{equation}
 \boxed{\;
 S_{\rm line}=\frac1\kappa\int dt\,d^3x
 \left[
 \operatorname{Tr}(\Pi\,\dot L L^{-1})
 +\zeta^i{\cal G}_i-s^i{\cal V}_i+\omega{\cal H}
 \right].
 \;}
 \label{eq:lineaction}
\end{equation}

Variation of $(\zeta^i,s^i,\omega)$ gives
\eqref{eq:sevenconstraints}.
For the remaining Euler--Lagrange equations use the $U$ defined
above.  Define the two bulk functional gradients
${\cal E}_\Pi$ and ${\cal E}_L$ by
\begin{align}
 {\cal C}&:=\int_\Sigma
 (\zeta^i{\cal G}_i-s^i{\cal V}_i+\omega{\cal H}),
 \nonumber\\
 \delta{\cal C}&=\int_\Sigma
 \left[
 \operatorname{Tr}({\cal E}_\Pi\,\delta\Pi)
 +\operatorname{Tr}({\cal E}_L\,\delta L L^{-1})
 \right],
 \label{eq:gradients}
\end{align}
with boundary and multiplier-variation terms understood.
The equations of motion are then
\begin{equation}
 \boxed{\qquad
 U=-{\cal E}_\Pi,\qquad
 \dot\Pi=[U,\Pi]+{\cal E}_L .
 \qquad}
 \label{eq:ELeq}
\end{equation}

These are the Pleba\'nski field equations rewritten entirely in line
variables.  Let $\psi^{ij}$ denote the null-basis form of the
original Pleba\'nski multiplier $\Psi^{IJ}$; it is symmetric and
$h$-tracefree.  Then
\begin{equation}
 {\cal R}^{ij}=D(\psi^{ij}+\Lambda h^{ij}),\qquad
 {\cal W}^{ij}:=D^{-1}{\cal R}^{ij}-\Lambda h^{ij}
 =\psi^{ij}.
 \label{eq:Weylidentity}
\end{equation}
The relation with the usual spacetime curvature is now transparent.
The simple two-forms $B_i$ determine the Urbantke metric and span one
of the two three-dimensional chiral subspaces of complexified
two-forms.  The connection equation obtained by varying the
Pleba\'nski connection makes that connection the corresponding
chiral Levi--Civita connection of the Urbantke metric.  Consequently
$D^{-1}{\cal R}^{ij}$ is the matrix, in the null basis $B_i$, of the
chiral part of the spacetime Riemann curvature.

For the Levi--Civita curvature, the usual decomposition of the
Riemann tensor with respect to the self-dual and anti-self-dual
two-form subspaces separates three pieces: the Weyl curvature in the
two diagonal chiral blocks, the scalar-curvature contribution in
their traces, and the tracefree Ricci tensor in the off-diagonal
blocks.  The Pleba\'nski field equations set the tracefree Ricci
part to zero.  In the present chiral block the trace is therefore
the cosmological-curvature term $\Lambda h^{ij}$, while the
symmetric $h$-tracefree remainder
\[
 {\cal W}^{ij}
 =D^{-1}{\cal R}^{ij}-\Lambda h^{ij}
\]
is precisely the self-dual Weyl curvature.  Thus ${\cal R}^{ij}$
is not the Ricci tensor: it is the chiral curvature matrix.  The
Einstein equations make the spacetime Ricci tensor pure trace
(with normalization fixed by the convention for $\Lambda$ in the
action), whereas the five independent components of
${\cal W}^{ij}$ contain the nontrivial self-dual Weyl tensor.

Hence
\eqref{eq:lineaction} is equivalent, on the nondegenerate simplicity
sector, to Einstein's equations with cosmological constant.

\subsection{Example: Pleba\'nski's second heavenly metric in three lines}
\label{subsec:second-heavenly}

The preceding chiral interpretation becomes especially transparent
for Pleba\'nski's second heavenly metrics~\cite{Plebanski1975}.
Let $\Theta=\Theta(p,q,x,y)$ be the heavenly potential; here
$\Theta$ is unrelated to the symplectic potential
$\Theta_{\rm line}$.  The second heavenly equation is
\begin{equation}
 \boxed{
 \Theta_{xx}\Theta_{yy}-\Theta_{xy}^{\,2}
 +\Theta_{xp}+\Theta_{yq}=0 .
 }
 \label{eq:second-heavenly}
\end{equation}
It determines the complex Ricci-flat symmetric bilinear form
$g_\Theta$ on the coordinate basis $(dp,dq,dx,dy)$, defined by
\begin{equation}
 \begin{aligned}
 g_\Theta:={}&
 dp\otimes dx+dx\otimes dp
 +dq\otimes dy+dy\otimes dq
 \\
 &-2\Theta_{yy}\,dp\otimes dp
 -2\Theta_{xx}\,dq\otimes dq
 \\
 &+2\Theta_{xy}
 \bigl(dp\otimes dq+dq\otimes dp\bigr).
 \end{aligned}
 \label{eq:second-heavenly-metric-tensor}
\end{equation}
Equivalently, in the quadratic-form notation
\eqref{eq:Gtods}, its line element is
\begin{equation}
 \boxed{
 \begin{aligned}
 ds_\Theta^2={}&
 2\,dp\left(dx-\Theta_{yy}\,dp+\Theta_{xy}\,dq\right)\\
 &+2\,dq\left(dy+\Theta_{xy}\,dp-\Theta_{xx}\,dq\right).
 \end{aligned}}
 \label{eq:second-heavenly-metric}
\end{equation}
Define the four heavenly vector fields
\begin{equation}
 \begin{aligned}
 X_{\rm h}&:=\partial_x,\qquad
 Y_{\rm h}:=\partial_y,\\
 U_{\rm h}&:=\partial_p
 +\Theta_{yy}X_{\rm h}-\Theta_{xy}Y_{\rm h},\\
 V_{\rm h}&:=\partial_q
 -\Theta_{xy}X_{\rm h}+\Theta_{xx}Y_{\rm h}.
 \end{aligned}
 \label{eq:heavenly-frame}
\end{equation}
At each point choose the affine splitting
\begin{equation}
 \begin{aligned}
 T_{\mathbb C}M&=\mathbb C e_0\oplus E_{\rm h},\\
 e_0&:=X_{\rm h},\qquad
 E_{\rm h}:=
 \operatorname{span}_{\mathbb C}
 \{Y_{\rm h},V_{\rm h},U_{\rm h}-X_{\rm h}\}.
 \end{aligned}
 \label{eq:heavenly-affine-splitting}
\end{equation}
Then the heavenly geometry is represented by the three affine lines
in $E_{\rm h}$
\begin{equation}
 \boxed{
 \begin{aligned}
 \ell_1&=\mathbb C\,Y_{\rm h},\\
 \ell_2&=(U_{\rm h}-X_{\rm h})+\mathbb C\,V_{\rm h},\\
 \ell_3&=(U_{\rm h}-X_{\rm h}-Y_{\rm h})
 +\mathbb C\,
 (V_{\rm h}-U_{\rm h}+X_{\rm h}+Y_{\rm h}).
 \end{aligned}}
 \label{eq:heavenly-lines}
\end{equation}
A convenient common scaling of their Pl\"ucker representatives is
\begin{equation}
 \boxed{
 \begin{aligned}
 \cP_1&=2X_{\rm h}\wedge Y_{\rm h},\qquad
 \cP_2=2U_{\rm h}\wedge V_{\rm h},\\
 \cP_3&=
 2(U_{\rm h}-Y_{\rm h})\wedge(V_{\rm h}+X_{\rm h}).
 \end{aligned}}
 \label{eq:heavenly-Plucker}
\end{equation}
Indeed,
\[
 (e_0+U_{\rm h}-X_{\rm h}-Y_{\rm h})
 \wedge
 (V_{\rm h}-U_{\rm h}+X_{\rm h}+Y_{\rm h})
 =
 (U_{\rm h}-Y_{\rm h})\wedge(V_{\rm h}+X_{\rm h}).
\]
Choose
\begin{equation}
 \Omega:=dp\wedge dq\wedge dx\wedge dy,
 \qquad
 B_i:=\iota_{\cP_i}\Omega .
 \label{eq:heavenly-B}
\end{equation}
On the heavenly equation \eqref{eq:second-heavenly} these forms obey
\begin{equation}
 \boxed{\begin{aligned}
& B_i\wedge B_i=0
 \quad (i=1,2,3;\ \text{no sum over $i$}),\\
& B_1\wedge B_2
 =B_1\wedge B_3
 =B_2\wedge B_3
 =4\,\Omega .\end{aligned}
 }
 \label{eq:heavenly-simplicity}
\end{equation}
Thus they realize precisely the null-basis simplicity relation
\eqref{eq:simplicity}, with ${\cal V}_4=4\Omega$.

Substitution of \eqref{eq:heavenly-Plucker} into the direct
three-line Urbantke formula \eqref{eq:urb} gives
\begin{equation}
 \boxed{
 \cG(\,\cdot\,,\,\cdot\,)
 =4\,g_\Theta(\,\cdot\,,\,\cdot\,)\,\Omega ,
 }
 \label{eq:heavenly-Urbantke}
\end{equation}
so that the ordinary metric representative defined by
\eqref{eq:Gtog} is exactly
\begin{equation}
 g=4g_\Theta .
 \label{eq:heavenly-factor-four}
\end{equation}
The factor $4$ is conformally irrelevant but fixes the normalization
of the chosen Pl\"ucker representatives.

Let $*$ denote the Hodge operator on two-forms defined by
$g_\Theta$ and the orientation determined by
$\Omega=dp\wedge dq\wedge dx\wedge dy$.

For the orientation in \eqref{eq:heavenly-B}, the three forms are
self-dual,
\begin{equation}
 *B_i=B_i , \qquad i=1,2,3,
 \label{eq:heavenly-selfdual}
\end{equation}
and the heavenly equation also implies
\begin{equation}
 dB_i=0 \qquad i=1,2,3.
 \label{eq:heavenly-closed}
\end{equation}
Hence the compatible connection in this selected chiral sector can
be chosen to vanish.  Its curvature therefore vanishes, and for
$\Lambda=0$ the line-curvature quantities introduced above satisfy
\begin{equation}
 \boxed{
 {\cal F}_{ij}{}^k=0,\qquad
 {\cal R}^{ij}=0,\qquad
 {\cal W}^{ij}=0 .
 }
 \label{eq:heavenly-line-Weyl-zero}
\end{equation}

This does not imply that the spacetime is flat: only the Weyl
chirality represented by the chosen $B_i$ vanishes.  For example,
\begin{equation}
 \Theta=\frac{x^4}{24}
 \label{eq:heavenly-witness}
\end{equation}
satisfies \eqref{eq:second-heavenly} and gives
\begin{equation}
 R_{\mu\nu}=0,
 \qquad
 R_{\mu\nu\rho\sigma}\not\equiv0 ,
 \label{eq:heavenly-nonflat}
\end{equation}
where $R_{\mu\nu}$ and $R_{\mu\nu\rho\sigma}$ denote respectively
the Ricci and Riemann tensors of $g_\Theta$. 
Thus the opposite chiral Weyl block is nonzero even though
\eqref{eq:heavenly-line-Weyl-zero} holds.  The example makes
explicit that ${\cal W}^{ij}$ is one chiral half of the spacetime
Weyl curvature rather than the full Weyl tensor.

\section{Canonical and constraint algebra}

Before eliminating the momentum, it is useful to display explicitly
the canonical algebra and the complete first-class algebra of the
seven constraints.  The symplectic potential
\eqref{eq:symplectic} gives the equal-time brackets
\begin{align}
 \{L_i{}^a(x),L_j{}^b(y)\}&=0,\nonumber\\
 \{L_i{}^a(x),\Pi_j{}^k(y)\}
 &=\kappa\,\delta_i{}^kL_j{}^a(x)\,\delta^3(x-y),
 \label{eq:canonicalLP}\\
 \{\Pi_i{}^j(x),\Pi_k{}^\ell(y)\}
 &=\kappa\left(
 \delta_i{}^\ell\Pi_k{}^j
 -\delta_k{}^j\Pi_i{}^\ell
 \right)(x)\,\delta^3(x-y).
 \label{eq:canonicalPiPi}
\end{align}
Here $\delta^3(x-y)$ denotes the coordinate Dirac distribution on
$\Sigma$.  Thus the momentum variables realize pointwise the
$\mathfrak{gl}(3,\C)$ algebra associated with the right-trivialized
cotangent bundle $T^*GL(3,\C)$.

For a weight-zero internal smearing $\phi^i$, define the smeared
Gauss generator
\begin{equation}
 \mathsf G[\phi]
 :=\frac1\kappa\int_\Sigma\phi^i{\cal G}_i .
 \label{eq:smearedG}
\end{equation}
For an ordinary spatial vector field $N^a$, define its line-frame
components by
\[
 \mathfrak s^i[N]:=N^aR_a{}^i ,
 \qquad
 N^a=\mathfrak s^i[N]L_i{}^a ,
\]
and define the raw vector generator
\begin{equation}
 \mathsf C[N]
 :=\frac1\kappa\int_\Sigma
 \mathfrak s^i[N]\,{\cal V}_i .
 \label{eq:smearedC}
\end{equation}

The raw vector generator differs from the generator of an ordinary
spatial diffeomorphism by a field-dependent internal rotation.
Using
\[
 N^aa_a{}^i=-N^aR_a{}^j\Pi_j{}^i ,
\]
define the improved spatial-diffeomorphism generator by
\begin{equation}
 \mathsf D[N]
 :=\mathsf C[N]-\mathsf G[N^aa_a].
 \label{eq:improvedD}
\end{equation}
Here $N^aa_a$ denotes the internal smearing with components
$(N^aa_a)^i:=N^aa_a{}^i$.  The generator $\mathsf D[N]$ produces
the ordinary spatial Lie derivative on the canonical fields.

For a scalar density $u$ of weight $-1$, define
\begin{equation}
 \mathsf H[u]
 :=\frac1\kappa\int_\Sigma u\,{\cal H}.
 \label{eq:smearedH}
\end{equation}
For closed $\Sigma$, or for compactly supported smearings, these
generators satisfy the complete first-class algebra
\begin{align}
 \{\mathsf G[\phi],\mathsf G[\varphi]\}
 &=\mathsf G[[\phi,\varphi]],\nonumber\\
 \{\mathsf D[N],\mathsf G[\phi]\}
 &=\mathsf G[{\cal L}_N\phi],\nonumber\\
 \{\mathsf D[N],\mathsf D[N']\}
 &=\mathsf D[[N,N']],\nonumber\\
 \{\mathsf G[\phi],\mathsf H[u]\}
 &=0,\nonumber\\
 \{\mathsf D[N],\mathsf H[u]\}
 &=\mathsf H[{\cal L}_Nu].
 \label{eq:constraint-algebra}
\end{align}
The brackets of smearings appearing here are
\begin{align}
 [\phi,\varphi]^k
 &:=f_{ij}{}^k\phi^i\varphi^j,
 \qquad
 f_{ij}{}^k\ \hbox{as in \eqref{eq:nullstructure}},
 \nonumber\\
 [N,N']^a
 &:=N^b\partial_bN'^a-N'^b\partial_bN^a .
 \label{eq:smearing-brackets}
\end{align}
The Lie derivatives of the smearings are
\begin{equation}
 ({\cal L}_N\phi)^i:=N^a\partial_a\phi^i,
 \qquad
 {\cal L}_Nu:=N^a\partial_au-u\,\partial_aN^a .
 \label{eq:Lie-eta}
\end{equation}

The Hamiltonian--Hamiltonian bracket contains a field-dependent
structure function.  Define the symmetric density
\begin{equation}
 \mathsf Q^{ab}:=h^{ij}L_i{}^aL_j{}^b
 \label{eq:Qstructure}
\end{equation}
and, for two weight-$-1$ scalar densities $u,v$, define the spatial
vector field
\begin{equation}
 W^a(u,v)
 :=2\mathsf Q^{ab}
 \left(u\,\partial_bv-v\,\partial_bu\right).
 \label{eq:Sstructure}
\end{equation}
Then
\begin{equation}
 \boxed{
 \{\mathsf H[u],\mathsf H[v]\}
 =-\mathsf C[W]
 =-\mathsf D[W]-\mathsf G[W^aa_a].
 }
 \label{eq:HHconstraint}
\end{equation}
In particular, the cosmological term does not modify this bracket.
Equations~\eqref{eq:constraint-algebra} and
\eqref{eq:HHconstraint} show explicitly that the seven line
constraints form a first-class system.

\subsection{Two-surface area in line variables}

As a first geometric observable, consider an oriented two-surface
$S\subset\Sigma$ in a real spacelike sector of the theory.  Let
$(\sigma^1,\sigma^2)$ be local coordinates on $S$ and
$x^a=x^a(\sigma^1,\sigma^2)$ its embedding.  Define its densitized
conormal by
\begin{equation}
 \mathfrak n_a
 :=\widetilde\epsilon_{abc}
 \frac{\partial x^b}{\partial\sigma^1}
 \frac{\partial x^c}{\partial\sigma^2},
 \qquad
 \widetilde\epsilon_{123}=+1 .
 \label{eq:surface-conormal}
\end{equation}
Here $\widetilde\epsilon_{abc}$ is the numerical alternating spatial
density in the chosen oriented coordinate chart.  The three line
fluxes through $S$ are
\begin{equation}
 {\mathfrak F}_i:=L_i{}^a\mathfrak n_a .
 \label{eq:line-fluxes}
\end{equation}
In terms of these quantities the ordinary geometric area of $S$ is
\begin{equation}
 \boxed{
 A[S;L]
 =\int_S
 \sqrt{h^{ij}{\mathfrak F}_i{\mathfrak F}_j}\,
 d^2\sigma .
 }
 \label{eq:line-area}
\end{equation}
where $d^2\sigma:=d\sigma^1d\sigma^2$.  Thus the area depends only
on the line configuration $L$ and not on its conjugate momentum
$\Pi$.

For the null-basis metric used above,
\[
 h^{ij}
 =\frac12
 \begin{pmatrix}
 -1&1&1\\
 1&-1&1\\
 1&1&-1
 \end{pmatrix},
\]
the pointwise area density has the Heron-type form
\begin{equation}
 h^{ij}{\mathfrak F}_i{\mathfrak F}_j
 =\frac12\left(
 -{\mathfrak F}_1^2-{\mathfrak F}_2^2-{\mathfrak F}_3^2
 +2{\mathfrak F}_1{\mathfrak F}_2
 +2{\mathfrak F}_1{\mathfrak F}_3
 +2{\mathfrak F}_2{\mathfrak F}_3
 \right).
 \label{eq:Heron-line-area}
\end{equation}
The branch of the square root is fixed by the chosen real spacelike
sector.

Let $\Phi[L]$ denote a formal kinematical wave functional in the
$L$-representation.  The corresponding formal area operator acts
by multiplication:
\begin{equation}
 \boxed{
 \widehat A[S]\Phi[L]
 =A[S;L]\,\Phi[L].
 }
 \label{eq:area-operator}
\end{equation}
No discreteness of its kinematical spectrum follows from this
representation alone.

\section{Elimination of the line momentum and the second-order
Lagrangian}
\label{sec:momentum-elimination}

Although the curvature \eqref{eq:linecurvature} contains spatial
derivatives of $\Pi$, these derivatives enter the action linearly.
After the corresponding spatial integration by parts, the
Euler--Lagrange equation obtained by varying $\Pi$ contains no
spatial derivatives of $\Pi$.  Thus the Legendre equation for
$\Pi$ is pointwise algebraic.  This is the structural reason why
the momentum can be eliminated explicitly on the regular Legendre
sector defined below.

Define
\begin{align}
 {\mathsf M}_{ij}&:=\epsilon_{kij}s^k+\omega h_{ij},
 \qquad
 {\mathsf M}_i{}^j:={\mathsf M}_{ik}h^{kj},
 \nonumber\\
 \chi&:=\omega^2+\frac12h_{ij}s^is^j .
 \label{eq:Mleg}
\end{align}
so that $\det{\mathsf M}=\omega\chi$.  All matrix products and
traces involving $\mathsf M$ refer to the mixed endomorphism
${\mathsf M}_i{}^j$.

Define
\begin{align}
 {\mathsf S}_r{}^{ij}
 &:={\mathsf M}_{kr}\epsilon^{kij},
 \nonumber\\
 {\mathsf B}_r{}^j
 &:=
 -\zeta^m f_{mr}{}^j
 +\nabla_i^{(-1)}{\mathsf S}_r{}^{ij}
 +\gamma_i{\mathsf S}_r{}^{ij}
 +\frac12{\mathsf S}_r{}^{ik}\Gamma_{ik}{}^j,
 \nonumber\\
 {\mathsf Y}&:=U+{\mathsf B}.
 \label{eq:Baff}
\end{align}

Let $\mathbf 1_3$ denote the $3\times3$ identity matrix.  The first
Hamilton equation can be written
\[
 {\mathsf Y}=-\mathbb L(\Pi),
\]
where the local linear Legendre operator
\[
 \mathbb L:
 \operatorname{Mat}_{3\times3}(\C)
 \longrightarrow
 \operatorname{Mat}_{3\times3}(\C)
\]
is
\begin{equation}
 \boxed{
 \begin{aligned}
 \mathbb L(X)=\sqrt2\Big[
 &X{\mathsf M}+{\mathsf M}X
 -(\operatorname{Tr}{\mathsf M})X
 -(\operatorname{Tr}X){\mathsf M}\\
 &+\bigl(
 (\operatorname{Tr}{\mathsf M})(\operatorname{Tr}X)
 -\operatorname{Tr}({\mathsf M}X)
 \bigr)\mathbf 1_3
 \Big].
 \end{aligned}}
 \label{eq:Legendre-operator}
\end{equation}
Equivalently,
\begin{equation}
 \boxed{
 \mathbb L=\sqrt2\,D(\operatorname{adj})_{\mathsf M},
 }
 \label{eq:Legendre-adjugate}
\end{equation}
where $\operatorname{adj}$ is the adjugate map, characterized by
\[
 {\mathsf M}\operatorname{adj}({\mathsf M})
 =
 \operatorname{adj}({\mathsf M}){\mathsf M}
 =
 (\det{\mathsf M})\mathbf 1_3 .
\]

On the regular sector
$\det{\mathsf M}=\omega\chi\neq0$, the inverse operator is
\begin{equation}
 \boxed{
 \mathbb L^{-1}(Z)
 =
 \frac{
 \frac12\operatorname{Tr}(Z{\mathsf M}){\mathsf M}
 -{\mathsf M}Z{\mathsf M}}
 {\sqrt2\,\det{\mathsf M}} .
 }
 \label{eq:Legendre-inverse}
\end{equation}
Hence ${\mathsf Y}=-\mathbb L(\Pi)$ gives
\begin{equation}
 \boxed{
 \Pi=
 \frac{
 {\mathsf M}{\mathsf Y}{\mathsf M}
 -\frac12\operatorname{Tr}({\mathsf Y}{\mathsf M}){\mathsf M}}
 {\sqrt2\,\det{\mathsf M}} .
 }
 \label{eq:Pielem}
\end{equation}

Substitution in \eqref{eq:lineaction} gives
\begin{align}
 {\cal L}_{(2)}
 =\frac1\kappa\Bigg[
 &\zeta^i\gamma_i-3\Lambda\omega D \nonumber\\
 &+\frac{
 \operatorname{Tr}({\mathsf Y}{\mathsf M}{\mathsf Y}{\mathsf M})
 -\frac12[\operatorname{Tr}({\mathsf Y}{\mathsf M})]^2}
 {2\sqrt2\,\det{\mathsf M}}
 \Bigg].
 \label{eq:L2}
\end{align}
Its only dynamical configuration is the weighted triple of lines
$L$.

For the singular-Legendre analysis below it is convenient to isolate
the momentum-dependent Legendre term in the square brackets of
\eqref{eq:L2}.  Define
\begin{equation}
 \ell_{\rm Leg}
 :=
 \frac{
 \operatorname{Tr}({\mathsf Y}{\mathsf M}{\mathsf Y}{\mathsf M})
 -\frac12[\operatorname{Tr}({\mathsf Y}{\mathsf M})]^2}
 {2\sqrt2\,\det{\mathsf M}},
 \qquad
 \widehat{\mathcal L}:=\sqrt2\,\ell_{\rm Leg}.
 \label{eq:scaled-Legendre-density}
\end{equation}
Thus the hats used below refer to this scaled Legendre sector; the
momentum-independent terms
$\zeta^i\gamma_i-3\Lambda\omega D$ in \eqref{eq:L2} are not included
in $\widehat{\mathcal L}$.

Finally,
\begin{equation}
 \det\mathbb L
 =32\sqrt2\,\omega^3\chi^3,
 \label{eq:Legendre-determinant}
\end{equation}
so the first- and second-order formulations are locally equivalent
whenever $\omega\chi\neq0$.

\section{The singular Legendre surface $\chi=0$}
\label{sec:singular-Legendre}

The second zero of the Legendre determinant has a simple geometric
meaning.  A representative of the same Urbantke conformal metric,
written in the $3+1$ variables introduced above, has components
\begin{align}
 G_{00}&=2D^2\omega\chi,
 \nonumber\\
 G_{0a}&=D^2\omega\,R_a{}^i h_{ij}s^j,
 \nonumber\\
 G_{ab}&=D^2\omega\,R_a{}^iR_b{}^j h_{ij},
 \label{eq:metric-3plus1}
\end{align}
and
\begin{equation}
 \boxed{\det G=4D^6\omega^6.}
 \label{eq:metric-det}
\end{equation}
Thus $\omega=0$ is a genuine degeneration of the metric; indeed,
from \eqref{eq:dynamiclines} the three affine lines then become
concurrent through $-N$.  By contrast, if
\[
 D\omega\neq0,\qquad \chi=0,
\]
the metric remains nondegenerate while
\begin{equation}
 \boxed{G_{00}=G(\partial_t,\partial_t)=0.}
 \label{eq:null-evolution}
\end{equation}
Hence $\chi=0$ is not a singularity of the four-geometry: it is the
locus on which the chosen evolution vector $\partial_t$ becomes
null.  The singularity of \eqref{eq:L2} is therefore a singularity
of the fully reduced Legendre chart.

The algebraic origin of this degeneracy is now transparent from
\eqref{eq:Legendre-adjugate}.  For a spectral parameter $z\in\C$,
the characteristic polynomial of $\mathsf M$ is
\begin{equation}
 \det(z\mathbf 1_3-\mathsf M)
 =(z-\omega)(z^2-2\omega z+\chi).
 \label{eq:M-characteristic}
\end{equation}
Consequently, on $\chi=0$ with $\omega\neq0$,
\begin{equation}
 \operatorname{spec}(\mathsf M)=\{0,\omega,2\omega\},
 \qquad
 \operatorname{rank}\mathbb L=6.
 \label{eq:Legendre-rank-six}
\end{equation}
The nine-dimensional Legendre map therefore acquires a
three-dimensional kernel.

In an eigenframe adapted to the ordered eigenvalues
$(2\omega,\omega,0)$,
\begin{equation}
 \mathsf M_0=\operatorname{diag}(2\omega,\omega,0),
 \qquad \omega\neq0.
 \label{eq:M-eigenframe}
\end{equation}
Writing $\mathsf E_{ij}$ for the standard $3\times3$ matrix units,
a basis of
$\ker D(\operatorname{adj})_{\mathsf M_0}$ is
\begin{equation}
 \mathsf E_{12},\qquad
 \mathsf E_{21},\qquad
 \operatorname{diag}(-2,1,0).
 \label{eq:Legendre-kernel-basis}
\end{equation}
Therefore, if $\mathsf Y=(y_{ij})$, the equation
$\mathsf Y=-\mathbb L(\Pi)$ can be solved on $\chi=0$ only when
\begin{equation}
 \boxed{
 y_{12}=0,\qquad
 y_{21}=0,\qquad
 y_{22}-2y_{11}=0.
 }
 \label{eq:Legendre-compatibility}
\end{equation}
These are precisely the three compatibility conditions associated
with the three momentum components that cease to be reconstructible
from the velocity.

\subsection{Finite constrained completion}

The loss of rank does not require abandoning the second-order
description.  Instead one eliminates only the six momentum
directions on which the Legendre map remains invertible and retains
the remaining three kernel components as auxiliary variables.

Introduce a local spectral parameter $u$, of the same density weight
as $\omega$, and set
\begin{equation}
 \xi_2:=2\omega-u,\qquad
 \xi_1:=\omega,\qquad
 \xi_0:=u,\qquad
 \chi=u(2\omega-u).
 \label{eq:spectral-u}
\end{equation}
Thus $u=0$ is precisely the locus $\chi=0$ when $\omega\neq0$, and
the three eigenvalues of $\mathsf M$ in the corresponding local
eigenframe are $\xi_2,\xi_1,\xi_0$.

Write $\mathsf Y=(y_{ij})$ in this eigenframe and introduce three
auxiliary variables
$\upsilon_1,\upsilon_2,\upsilon_3$ for the momentum components that
become the kernel directions at $u=0$.  In the scaled normalization
\eqref{eq:scaled-Legendre-density}, the finite partially reduced
extension of $\widehat{\mathcal L}$ is
\begin{align}
 \widehat{\mathcal L}_{\rm reg}
 ={}&
 \frac{y_{13}y_{31}}{\xi_1}
 +\frac{y_{23}y_{32}}{\xi_2}
 -\frac{y_{11}y_{33}}{\xi_1}
 +\upsilon_1y_{21}+\upsilon_2y_{12}
 -u\,\upsilon_1\upsilon_2
 \nonumber\\
 &+\upsilon_3
 \left(
 \xi_1y_{22}-\xi_2y_{11}-u\,y_{33}
 \right)
 -\xi_2\xi_1u\,\upsilon_3^2 .
 \label{eq:finite-local-Lagrangian}
\end{align}
The auxiliary variables carry no time derivatives.  For $u\neq0$
their algebraic equations of motion determine
$\upsilon_1,\upsilon_2,\upsilon_3$, and eliminating them reproduces
exactly the scaled fully reduced Legendre density
$\widehat{\mathcal L}$.

At $u=0$, by contrast, the terms quadratic in the auxiliary
variables disappear and they become Lagrange multipliers.  Their
equations of motion reduce to
\begin{equation}
 y_{21}=0,\qquad
 y_{12}=0,\qquad
 y_{22}-2y_{11}=0,
 \label{eq:compatibility-from-auxiliary}
\end{equation}
which are precisely the compatibility conditions
\eqref{eq:Legendre-compatibility}.  Thus no propagating degree of
freedom has been added: the three auxiliary variables retain exactly
the momentum information which ceases to be determined by the
velocity when the Legendre map loses rank.

If desired, the relation between the transverse parameter $u$ and
the invariant quantity $\chi$ may itself be imposed by a scalar
multiplier $\varrho$ through the additional term
\begin{equation}
 \varrho\,[\,u(2\omega-u)-\chi\,].
 \label{eq:u-constraint}
\end{equation}
Since
\begin{equation}
 \left.
 \frac{\partial}{\partial u}
 \bigl[u(2\omega-u)\bigr]
 \right|_{u=0}
 =2\omega\neq0,
 \label{eq:u-transverse}
\end{equation}
$u$ is a regular local coordinate transverse to the surface
$\chi=0$.  The singular factor $1/\chi$ in the fully reduced
Lagrangian is therefore replaced by a finite constrained
description through the null-evolution surface.

\subsection{Invariant eigenframe-free completion}
\label{subsec:invariant-completion}

All linear operators in this subsection act pointwise on
$\operatorname{Mat}_{3\times3}(\C)$.  On $\chi=0$, with
$\omega\neq0$, the three eigenspaces of $\mathsf M$ can be selected
directly by the polynomial spectral projectors
\begin{equation}
\begin{aligned}
 P_0&=
 \frac{(\mathsf M-\omega\mathbf 1_3)
       (\mathsf M-2\omega\mathbf 1_3)}
      {2\omega^2},\\
 P_\omega&=
 \frac{\mathsf M(2\omega\mathbf 1_3-\mathsf M)}
      {\omega^2},
 \qquad
 P_{2\omega}=
 \frac{\mathsf M(\mathsf M-\omega\mathbf 1_3)}
      {2\omega^2}.
\end{aligned}
\label{eq:null-spectral-projectors}
\end{equation}
They are mutually orthogonal rank-one idempotents and satisfy
\[
 P_0+P_\omega+P_{2\omega}=\mathbf 1_3 .
\]

Set
\begin{equation}
 \mathsf Z:=P_\omega-2P_{2\omega},
 \qquad
 \operatorname{Tr}(\mathsf Z^2)=5 .
 \label{eq:kernel-Z}
\end{equation}
The exact rank-three projector onto the kernel of the degenerate
Legendre map is
\begin{equation}
 \boxed{
 K_{\mathsf M}(X)
 =
 P_{2\omega}XP_\omega
 +P_\omega XP_{2\omega}
 +\frac{\operatorname{Tr}(\mathsf Z X)}{5}\,\mathsf Z .
 }
 \label{eq:kernel-projector}
\end{equation}

For brevity set
\begin{equation}
 \mathfrak D_{\mathsf M}
 :=D(\operatorname{adj})_{\mathsf M}
 =\frac1{\sqrt2}\,\mathbb L .
 \label{eq:Dadj-short}
\end{equation}
Then
\begin{equation}
 K_{\mathsf M}^2=K_{\mathsf M},
 \qquad
 \operatorname{Im}K_{\mathsf M}
 =\ker\mathfrak D_{\mathsf M},
 \qquad
 \mathfrak D_{\mathsf M}K_{\mathsf M}
 =K_{\mathsf M}\mathfrak D_{\mathsf M}=0 .
 \label{eq:kernel-projector-properties}
\end{equation}
Consequently the three compatibility conditions
\eqref{eq:Legendre-compatibility} are equivalently the single
basis-independent equation
\begin{equation}
 \boxed{
 K_{\mathsf M}(\mathsf Y)=0 .
 }
 \label{eq:invariant-compatibility}
\end{equation}

The complementary six-dimensional sector admits the polynomial
group inverse
\begin{equation}
 \boxed{
 \mathbb G_{\mathsf M}
 =
 \frac1{\omega}
 \left[
 \frac{39}{20}\frac{\mathfrak D_{\mathsf M}}{\omega}
 +\frac34
  \left(\frac{\mathfrak D_{\mathsf M}}{\omega}\right)^2
 -\frac7{20}
  \left(\frac{\mathfrak D_{\mathsf M}}{\omega}\right)^3
 -\frac3{20}
  \left(\frac{\mathfrak D_{\mathsf M}}{\omega}\right)^4
 \right].
 }
 \label{eq:group-inverse}
\end{equation}
It obeys
\begin{equation}
 \mathfrak D_{\mathsf M}\mathbb G_{\mathsf M}
 =
 \mathbb G_{\mathsf M}\mathfrak D_{\mathsf M}
 =
 \mathbb I-K_{\mathsf M},
 \qquad
 K_{\mathsf M}\mathbb G_{\mathsf M}
 =
 \mathbb G_{\mathsf M}K_{\mathsf M}=0 ,
 \label{eq:group-inverse-properties}
\end{equation}
where $\mathbb I$ denotes the identity operator on
$\operatorname{Mat}_{3\times3}(\C)$.

Introduce a matrix-valued auxiliary field $\Upsilon$ constrained to
the kernel bundle,
\begin{equation}
 K_{\mathsf M}(\Upsilon)=\Upsilon .
 \label{eq:Upsilon-kernel}
\end{equation}
In terms of the scaled momentum
$\Pi_{\rm sc}:=\sqrt2\,\Pi$, its general value on the singular
surface is
\begin{equation}
 \boxed{
 \Pi_{\rm sc}
 =-\mathbb G_{\mathsf M}(\mathsf Y)+\Upsilon .
 }
 \label{eq:invariant-momentum-reconstruction}
\end{equation}
Thus $\mathbb G_{\mathsf M}$ reconstructs the six momentum
components determined by the velocity, while $\Upsilon$ retains
precisely the three components which the degenerate Legendre map
cannot determine.

In the normalization \eqref{eq:scaled-Legendre-density}, the finite
invariant density on $\chi=0$ is
\begin{equation}
 \boxed{
 \widehat{\mathcal L}_{\rm inv}
 =
 -\frac12
 \operatorname{Tr}\!\left[
 \mathsf Y\,\mathbb G_{\mathsf M}(\mathsf Y)
 \right]
 +\operatorname{Tr}(\Upsilon\mathsf Y),
 \qquad
 K_{\mathsf M}(\Upsilon)=\Upsilon .
 }
 \label{eq:invariant-null-Lagrangian}
\end{equation}
Variation with respect to $\Upsilon$ gives
\eqref{eq:invariant-compatibility}, whereas differentiation with
respect to $\mathsf Y$ reproduces
\eqref{eq:invariant-momentum-reconstruction}.  In the local
eigenframe this density reduces to
\eqref{eq:finite-local-Lagrangian} at $u=0$, after a regular
redefinition of the diagonal auxiliary variable.

The construction contains no eigenvectors.  Under an internal
change of basis
\[
 \mathsf M\longmapsto
 \mathsf T\,\mathsf M\,\mathsf T^{-1},
 \qquad
 \mathsf T\in GL(3,\C),
\]
the projectors, $K_{\mathsf M}$ and $\mathbb G_{\mathsf M}$ transform
by the corresponding similarity action, and
\eqref{eq:invariant-null-Lagrangian} is invariant.

\subsection{Application: a regular constrained second-order chart
through $\chi=0$}
\label{subsec:regular-chart}

The preceding construction lives intrinsically on the singular
surface.  It also admits a smooth extension to a full neighbourhood
of $\chi=0$.  Retain the spectral parametrization
\eqref{eq:spectral-u},
\[
 \xi_2=2\omega-u,\qquad
 \xi_1=\omega,\qquad
 \xi_0=u,\qquad
 \chi=u(2\omega-u).
\]
In a sufficiently small neighbourhood of $u=0$ the three eigenvalues
$\xi_2,\xi_1,\xi_0$ are distinct.  Their spectral projectors can be
written without eigenvectors as the Lagrange-interpolation
polynomials
\begin{align}
 P_{\xi_2}
 &=
 \frac{(\mathsf M-\xi_1\mathbf 1_3)
       (\mathsf M-\xi_0\mathbf 1_3)}
      {(\xi_2-\xi_1)(\xi_2-\xi_0)},
 \nonumber\\
 P_{\xi_1}
 &=
 \frac{(\mathsf M-\xi_2\mathbf 1_3)
       (\mathsf M-\xi_0\mathbf 1_3)}
      {(\xi_1-\xi_2)(\xi_1-\xi_0)},
 \nonumber\\
 P_{\xi_0}
 &=
 \frac{(\mathsf M-\xi_2\mathbf 1_3)
       (\mathsf M-\xi_1\mathbf 1_3)}
      {(\xi_0-\xi_2)(\xi_0-\xi_1)} .
 \label{eq:spectral-projectors-u}
\end{align}
Define
\begin{equation}
 \mathsf Z_u
 :=-\xi_2P_{\xi_2}
   +\xi_1P_{\xi_1}
   -\xi_0P_{\xi_0}
 \label{eq:Zu}
\end{equation}
and the rank-three operator
\begin{equation}
 \boxed{
 K_u(X)
 :=
 P_{\xi_2}XP_{\xi_1}
 +P_{\xi_1}XP_{\xi_2}
 +\frac{\operatorname{Tr}(\mathsf Z_uX)}
        {\operatorname{Tr}(\mathsf Z_u^2)}\,\mathsf Z_u .
 }
 \label{eq:Ku}
\end{equation}
It is an idempotent,
\[
 K_u^2=K_u,
\]
and satisfies
\begin{equation}
 \boxed{
 \lim_{u\to0}K_u=K_{\mathsf M}.
 }
 \label{eq:Ku-limit}
\end{equation}
For $u\neq0$, the image of $K_u$ is an auxiliary rank-three
subbundle rather than the kernel of $\mathfrak D_{\mathsf M}$,
since the full Legendre map is then invertible.

Define the complementary projector
\begin{equation}
 \mathbb P_u^\perp:=\mathbb I-K_u .
 \label{eq:Puperp}
\end{equation}
The operator
\begin{equation}
 \mathbb A_u
 :=
 \mathbb P_u^\perp\circ\mathfrak D_{\mathsf M}
 \circ\mathbb P_u^\perp+K_u
 \label{eq:Au}
\end{equation}
is invertible in a neighbourhood of $u=0$.  Define its inverse
restricted back to the complementary six-dimensional sector by
\begin{equation}
 \mathbb G_u
 :=
 \mathbb P_u^\perp\circ
 \mathbb A_u^{-1}\circ
 \mathbb P_u^\perp .
 \label{eq:Gu}
\end{equation}

The auxiliary matrix $\Upsilon$ is now required to satisfy
\begin{equation}
 K_u(\Upsilon)=\Upsilon .
 \label{eq:Upsilon-Ku}
\end{equation}
Introduce
\begin{equation}
 \mathsf J_u
 :=
 \mathbb P_u^\perp
 \left(
 \mathsf Y+\mathfrak D_{\mathsf M}(\Upsilon)
 \right).
 \label{eq:Ju}
\end{equation}
The invariant partial Legendre transform is
\begin{equation}
 \boxed{
 \widehat{\mathcal L}_{\rm reg}^{\rm inv}
 =
 \operatorname{Tr}(\Upsilon\mathsf Y)
 +\frac12\operatorname{Tr}\!\left[
 \Upsilon\,\mathfrak D_{\mathsf M}(\Upsilon)
 \right]
 -\frac12\operatorname{Tr}\!\left[
 \mathsf J_u\,\mathbb G_u(\mathsf J_u)
 \right].
 }
 \label{eq:regular-Schur-Lagrangian}
\end{equation}

This single density interpolates between the two descriptions.  For
$u\neq0$, eliminating $\Upsilon$ reproduces exactly the scaled fully
reduced Legendre density $\widehat{\mathcal L}$.  In the opposite
limit,
\begin{equation}
 \boxed{
 \lim_{u\to0}
 \widehat{\mathcal L}_{\rm reg}^{\rm inv}
 =
 \widehat{\mathcal L}_{\rm inv}.
 }
 \label{eq:regular-Schur-limit}
\end{equation}
The auxiliary equation simultaneously reduces to
$K_{\mathsf M}(\mathsf Y)=0$.  Thus the divergence of the $L$-only
expression at $\chi=0$ is removed not by inverting a singular
Hessian, but by performing the Legendre transform only on its
regular six-dimensional sector and retaining the three unresolved
momentum components as auxiliary variables.  The resulting
constrained second-order chart passes smoothly through the
null-evolution surface while the spacetime metric remains
nondegenerate.

\end{document}